# Tracking ultrafast change of multiterahertz broadband response functions in a photoexcited Dirac semimetal Cd$_3$As$_2$ thin film

*Natsuki Kanda[1,2], Yuta Murotani[1], Takuya Matsuda[1], Manik Goyal[3], Salva Salmani-Rezaie[3], Jun Yoshinobu[1], Susanne Stemmer[3], and Ryusuke Matsunaga[1,2]\**

[1]The Institute for Solid State Physics, The University of Tokyo, Kashiwa, Chiba 277-8581, Japan.
[2]PRESTO, Japan Science and Technology Agency, 4-1-8 Honcho Kawaguchi, Saitama 332-0012, Japan.
[3]Materials Department, University of California, Santa Barbara, California 93106-5050, USA.

*Email: matsunaga@issp.u-tokyo.ac.jp

**ABSTRACT.**

The electromagnetic response of Dirac semimetals in the infrared and terahertz frequency ranges is attracting growing interest for potential applications in optoelectronics and nonlinear optics. The interplay between the free-carrier response and interband transitions in the gapless, linear dispersion relation plays a key role in enabling novel functionalities. Here we investigate ultrafast dynamics in thin films of a photoexcited Dirac semimetal $Cd_3As_2$ by probing the broadband response functions as complex quantities in the multiterahertz region (10-45 THz, 40-180 meV, or 7-30 μm), which covers the crossover between the inter and intraband response. We resolve dynamics of the photoexcited nonthermal electrons which merge with originally existing carriers to form a single thermalized electron gas and how it is facilitated by high-density excitation. We also demonstrate that a large reduction of the refractive index by 80% dominates the nonequilibrium infrared response, which can be utilized for designing ultrafast switches in active optoelectronics.

**TEXT.**

Recently enormous research efforts have been devoted to investigating the novel properties of three-dimensional (3D) topological Dirac and Weyl semimetals that arise from topologically or symmetry protected relativistic electrons[1,2]. Nonlinear interaction with light is also highly intriguing because of its applicability to controlling the topological phase of matter by strong laser pulses[3-5]. In addition, their responses to light at room temperature are also drawing considerable attention for potential applications in nonlinear optics and optoelectronic devices, especially in mid-infrared (MIR), far-infrared, or terahertz (THz) frequency range because of their gapless band structure. Photocurrent generation from the near-infrared (NIR) to 10 μm is expected to realize high-speed broadband photodetectors[6-8]. Saturation of absorption for intense NIR and MIR light has been incorporated in mode-locked pulsed lasers as ultrafast modulators[9,10]. Highly nonlinear current from accelerated massless electrons realizes efficient frequency conversion in THz regime[11-13]. Tunable, perfect absorbers with semimetal-based metamaterials were also proposed for THz sensing and filters[14,15]. Low-frequency stimulated emission is also expected for use in an infrared gain medium in a similar way to graphene[16,17]. Compared to graphene, which consists of a single atomic monolayer, the large interaction volume with light, the feasibility of fabricating large-area thin films, and the robustness to the environment make 3D topological semimetals advantageous for these applications.

To thoroughly explore these novel functionalities, an in-depth understanding of the responses away from equilibrium is indispensable. Significant attention has focused on the ultrafast carrier dynamics of $Cd_3As_2$, a prototypical 3D Dirac semimetal with a large carrier mobility, high Fermi velocity, chemical stability, low Fermi level, and quasi-linear band character in a wide energy scale[18]. Pump-probe spectroscopy studies in $Cd_3As_2$ have been reported in the NIR[19], MIR[9,20-23] and at THz frequencies[11,24-26], but there are some controversies. For example, the relaxation dynamics is analyzed typically within the two-temperature model (TTM), which provides the time evolution of temperatures in the electron and lattice subsystems. However, whether one can define the electron temperature, *i.e.*, the time scale of establishing a hot Fermi-Dirac distribution from photoexcited nonthermal electrons, is an issue to be resolved. Prior time-resolved MIR experiments have been performed by probing the intensity of transmitted or reflected pulses and therefore require careful treatments to derive the complex response function in the frequency space. Information on the complex response functions is particularly important in semimetals, because in addition to absorption, the permittivity described by the real-part dielectric function, $\epsilon_1(\omega)$, is also essential in their infrared response. Figures 1(a) and 1(b) show the optical absorption spectrum $\sigma_1(\omega)$ (semi-log plot) and the real-part dielectric function $\epsilon_1(\omega)$ in $Cd_3As_2$, respectively (details are described below). Across the screened plasma frequency, located at 10

THz, metallic screening ($\epsilon_1 < 0$) changes to dielectric screening ($\epsilon_1 > 0$) with flipping its sign. When carriers are photoinjected into the semimetal, the plasma frequency is elevated, accompanied with large modulation of $\epsilon_1$ and refractive index $n$, which drastically alters the boundary condition of light for infrared interband transition. THz time-domain spectroscopy can directly probe the response function without Kramers-Kronig conversion, but only for the DC-limit response with limited time resolution of ~1 ps. Therefore, for a comprehensive understanding of the interplay between inter and intraband transitions, it is important to track ultrafast change of nonequilibrium response functions as complex quantities within a broad energy scale from several tens to a few hundreds of meV, i.e., multiterahertz frequency range from 10 to several tens of THz[27,28].

In this work, we investigated the ultrafast dynamics of a photoexcited $Cd_3As_2$ thin film by probing transmittance in the multiterahertz range (10-45 THz in frequency, 40-180 meV in energy, or 7-30 μm in wavelength). Broadband time-domain spectroscopy allows us to directly evaluate ultrafast changes of the response functions at the crossover region of inter and intraband transitions. After NIR photoexcitation, we resolve transient spectra of induced absorption and loss function, which are significantly deformed at the initial stage, clarifying the thermalization timescale of photoexcited electrons. We also elucidate a large suppression of the refractive index, which plays a dominant role in the infrared response of the photoexcited Dirac semimetal.

The sample is an epitaxially-grown, (112)-oriented $Cd_3As_2$ thin film with 140-nm thickness on an 800-μm-thick CdTe substrate[29-31]. Details are described in Supporting Information (SI) Note 1. Figure 1(c) shows the model calculation of the band structure of $Cd_3As_2$, which has two Dirac nodes near the Γ point along the [001] directions[32]. Note that the linear dispersion relation of electrons associated with the symmetry-protected Dirac nodes in $Cd_3As_2$ is limited to a small energy scale of meV, and that the dispersion relation in the multiterahertz scale at a few hundreds of meV is described by the massless-like Kane model[33]. The linear response function of the film was evaluated by THz time-domain spectroscopy and Fourier-transform infrared spectroscopy (See further details in SI Note 3). All experiments in this work were performed at room temperature. The thick solid curves in Figs. 1(a) and 1(b) show the experimental results of the conductivity spectrum $\sigma_1(\omega)$ and dielectric function $\epsilon_1(\omega)$, respectively. The red curve shows an interpolated fit by assuming the Drude-type response (green) and the interband transition linear to frequency with the step-like onset at $2E_F$ (blue). The Fermi energy is estimated to be as small as $E_F$=58 meV, which is important to observe intrinsic dynamics of carriers in $Cd_2As_2$. We also checked the polarization dependence and concluded that in-plane anisotropy is negligibly small in our frequency window at room temperature owing to the quasi-3-fold rotational symmetry of the (112) surface.

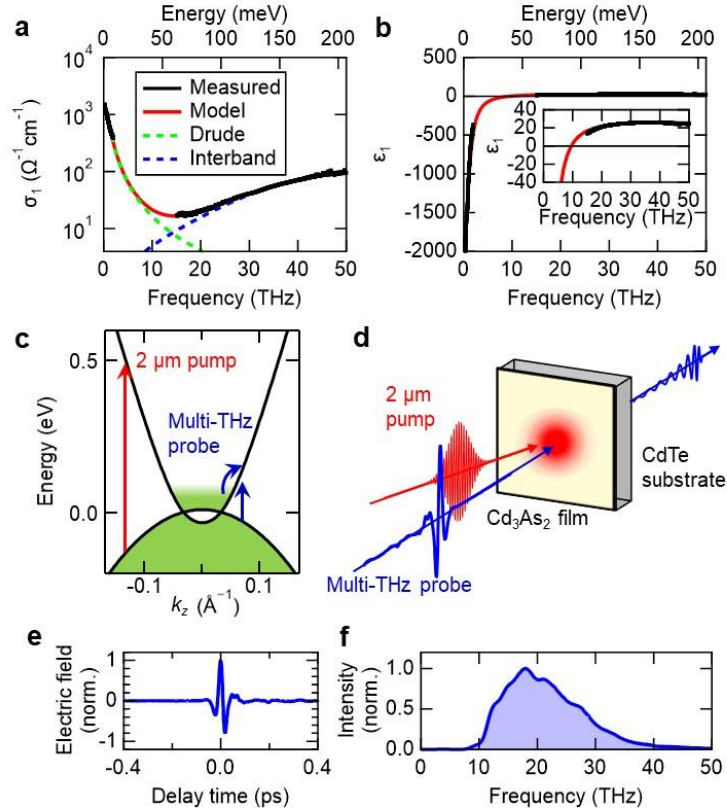

**Figure 1.** Equilibrium response function in Cd$_3$As$_2$ and experimental setup. (a)(b) Real-part optical conductivity and dielectric function in Cd$_3$As$_2$ in equilibrium at room temperature. The thick black curves show the experimental data, and the red curves show fitted results. The green and blue curves show the weights of the Drude response and interband transitions, respectively. The inset in (b) is enlarged data to emphasize the zero-crossing point at the screened plasma frequency of 10 THz. (c) Model calculation of the band dispersion in Cd$_3$As$_2$. The red and blue arrows indicate NIR pump and multiterahertz probe, respectively. (d) A schematic of our experimental setup. (e)(f) Waveform and power spectrum of the multiterahertz probe pulse.

To explore ultrafast dynamics after photoexcitation, we conducted pump-probe spectroscopy. Figure 1(d) shows a schematic of the experimental setup. The light source is a Yb:KGW-based regenerative amplifier with a center frequency of 1030 nm, a repetition rate of 3 kHz, a pulse energy of 2 mJ, and a pulse width of 255 fs. A portion of the output was compressed to <14 fs by using the multi-plate broadening scheme[34,35] and was further split to generate multiterahertz pulses and gate pulses. See details in SI Note 4. Figures 1(e) and 1(f) show the waveform and the power spectrum of multiterahertz pulses, which covers the broad bandwidth of 10-45 THz with the pulse duration of 28 fs. After transmission through the sample, the multiterahertz pulses were detected by electro-optic sampling. A residual laser output was sent to an optical parametric amplifier system to generate pump pulses with 2-μm wavelength (0.62 eV, 150 THz) and 230-fs pulse duration (see SI for further details).

Figure 2(a) shows the two-dimensional (2D) plot of differential transmittance $\Delta T/T$ as functions of frequency (horizontal) and pump delay (vertical). The pump fluence is 25 μJ/cm$^2$, which we refer to a weak excitation condition (see SI Note 6 for the detail). If one assumes that a Fermi-Dirac electron distribution is developed after photoexcitation with an elevated electron temperature, the TTM analysis gives a maximum electron temperature of 330 K for this pump fluence (see SI Note 2). The red- and blue-colored data show that $\Delta T$ tends to be positive at higher frequencies and negative at lower frequencies, respectively. Figure 2(b) shows the 2D plot of differential absorption spectra, *i.e.*, the change of the real-part conductivity $\Delta \sigma_1$. The red and blue colors correspond to bleaching ($\Delta \sigma_1 <0$) and induced absorption ($\Delta \sigma_1 >0$), respectively. Sequential spectral profiles of $\Delta \sigma_1$ at several pump delays are also plotted in Fig. 2(c). The green color indicates the timing of pump pulse irradiation. Right after the pump, induced absorption appears, which can be ascribed to absorption by photoexcited carriers. Notably, the induced absorption spectrum at 0.2 ps spreads over a wide frequency range, which cannot be attributed to be solely due to heating of electrons to the elevated temperature of 330 K. Subsequently, the spectral weight concentrates into the lower-frequency side to form a Drude-type spectrum, which occurs at the time scale of 0.5 ps. Simultaneously, the bleaching is also discerned at the higher frequency above 25 THz (~100 meV), which can be ascribed to the Pauli blocking in the interband transition. After that, the overall signal starts to decay while keeping its spectral shape. Note that $\Delta \sigma_1$ spectra in Fig. 2(c) show small dips at 13 and 25 THz. These features always appear only after the pump, which can be attributed to an artifact due to small absorptions of the substrate slightly modulated by the pump (SI Note 4 and Fig. S7).

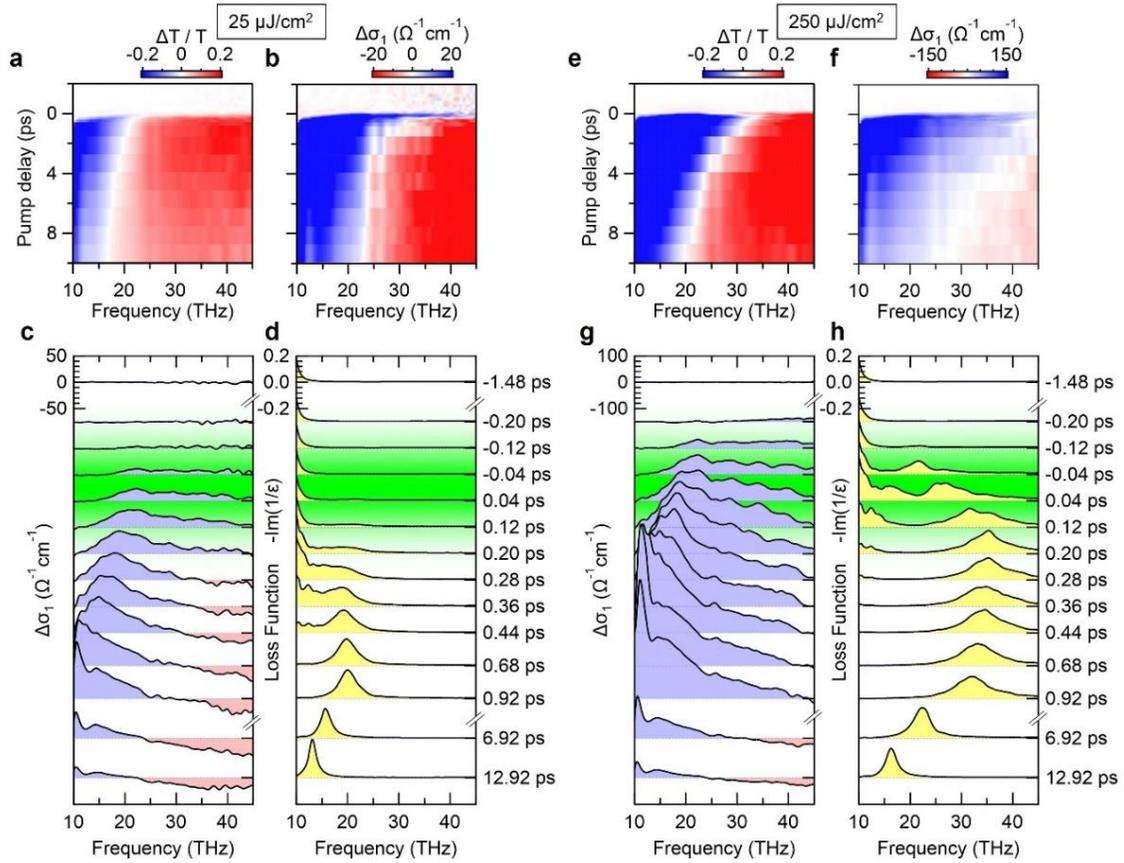

**Figure 2.** Ultrafast change of conductivity and loss function spectra. (a)(b) 2D plots of differential transmittance $\Delta T/T$ and differential conductivity spectrum $\Delta\sigma_1$ as functions of frequency (horizontal) and pump delay (vertical) for the weak pump excitation regime of 25 μJ/cm$^2$. (c) $\Delta\sigma_1$ at each pump delay. The green-colored region indicates the timing of the pump pulse irradiation. (d) Loss function spectrum $-\text{Im}(\epsilon^{-1})$ at each pump delay. (e)-(h) The same sets of data for the strong excitation regime of 250 μJ/cm$^2$.

We also plot the time evolution of $-\text{Im}(\epsilon^{-1})$ in Fig. 2(d). This is the loss function in the small wavevector limit and gives a peak of the longitudinal plasmon mode at the screened plasma frequency[36], which is located at 10 THz in equilibrium. Right after the photoexcitation, the loss function is significantly deformed and spread into higher frequency while leaving a spectral weight in the low frequency. The deformed spectrum indicates that the response is not well described by a simple Drude model, suggesting that photoexcited electrons in the initial stage are quite nonthermal and behave differently from originally existing electrons. After undergoing repeated scattering, they should be thermalized to form a single hot electron gas with a well-defined electron temperature. Indeed, the experimental result in Fig. 2(d) after ~0.5 ps shows that a single peak develops at 20 THz, indicating that a single collective plasmon mode becomes well defined at this time scale with an elevated plasma frequency. The result clarifies that the thermalization timescale of photoexcited electrons is as long as 0.5 ps. It is consistent with previous pump-probe studies[19,21] where a decay component of ~0.5 ps observed only in degenerate pump and probe energies was attributed to the thermalization process.

Figures 2(e)-2(h) show the data with a stronger excitation density of 250 µJ/cm$^2$, which corresponds to the maximum electron temperature of 730 K (see SI Note 2 and 7). Compared to the results in the weak excitation regime in Fig. 2(d), the loss function peak in Fig. 2(h) develops at 35 THz because of increased carrier density. In addition, the peak is significantly broadened, indicating that the scattering rate between the high-density carriers is enhanced. Correspondingly, the time scale of developing the single loss function peak is as short as 0.3 ps after the photoexcitation, which is faster than that in the weak excitation regime. It can be reasonably explained by the enhanced scattering between a high density of carriers, which facilitates thermalization in the electron system.

Next, we focus on the dynamics after thermalization. Figures 3(a) and 3(b) show the results of the real- and imaginary-part conductivity spectra $\sigma_1$ and $\sigma_2$ with the pump fluence of 50 µJ/cm$^2$ at various delays. The data are fitted well by thin curves assuming the Drude-type response and the reduced interband transition, which gives the dynamics of total carrier density as plotted in Fig. 3(c). In the weak excitation regime, the decay is well fitted by a single-exponential function. For higher excitation density >100 μJ/cm$^2$, a double-exponential function is required to fit a fast decay component (see SI Note 5). Figure 3(d) shows the decay constants as a function of the pump fluence. The slower decay constant $\tau_{\text{slow}}$ is around 10 ps in the weak-excitation limit and shortened to be ~8 ps in the high-density excitation. The results are consistent with or a bit longer than previous reports[9,11,19-26] and can be attributed to the carrier relaxation time interacting with the lattice system described by TTM. The fast decay component $\tau_{\text{fast}}$, observed only in the high-density excitation regime, is as fast as few hundred fs and not fully resolved due to the time resolution of the pump pulse. The fast decay may be attributed to Auger recombination, *i.e.*, electron-electron scattering decreases the carrier density. Figure 3(e) shows the total carrier density as a function of the pump fluence, and the weights of the fast- and slow-decay components were separately indicated. The result shows that the fast decay appears only when the total carrier density reaches to ~10$^{19}$ cm$^{-3}$, which corresponds to ~10$^{14}$ cm$^{-2}$ in terms of the sheet carrier density of the thin film. It is in a stark contrast to graphene, where the Auger recombination dominates the carrier relaxation due to efficient collinear scattering[37,38] even for much smaller sheet carrier density[39,40]. A recent theory shows that the Auger scattering is prohibited in 3D linear dispersions because of nondivergent matrix element in contrast to the 2D counterpart[41]. The electron dispersion relation in Cd$_3$As$_2$ in the multiterahertz energy scale is described by the Kane model with a nonvanishing parabolicity, which would allow to some degree of Auger scattering for high-density excitation and facilitate the relaxation as well as thermalization.

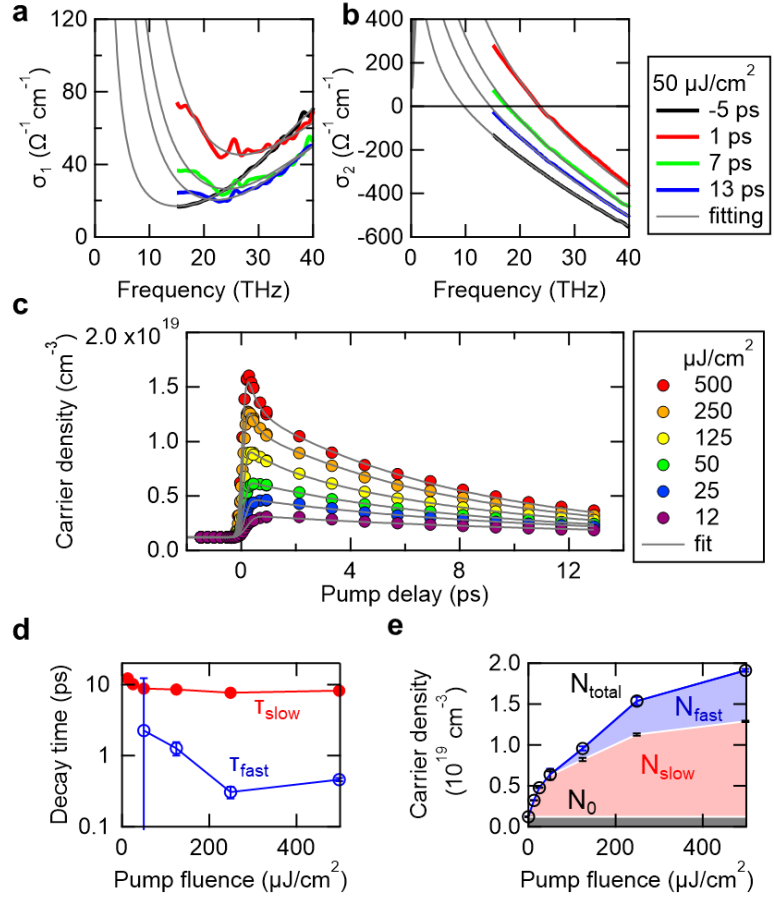

**Figure 3.** Dynamics after thermalization (>1 ps). (a)(b) Real and imaginary parts of the conductivity spectra, $\sigma_1$ and $\sigma_2$, for the excitation density of 50 μJ/cm$^2$ at various pump delays. Thin black curves show fitted results using the Drude-type response and interband transition. (c) Dynamics of total excited carrier density obtained by the fitting as a function of the pump delay. The fitting errors are smaller than the marker size. The data are fitted by single and double exponential functions below and above 50 μJ/cm$^2$, respectively. (d) Semi-log plot of the decay times obtained in the fitting in (c) as a function of the pump fluence. (e) Total carrier density at the maximum $N_{total}$ as a function of the pump fluence. The blue- red- and gray-colored areas ($N_{fast}, N_{slow}, N_0$) correspond to the weights of fast and slow decaying components and originally doped electrons, respectively.

Here we focus on the Pauli blocking signal ($\Delta\sigma_1<0$), *i.e.*, photoexcited electrons occupying the conduction band that prohibit further interband transitions. One could expect that, under strong pump, $\Delta\sigma_1<0$ could be so significant that $\sigma_1$ might flip its sign to negative at the onset of the interband transition, which is the stimulated emission by inverted population[16]. We examined this point with various pump fluences but have not identified the $\sigma_1<0$ signal in the present experimental conditions. The difficulty of the stimulated emission in the multiterahertz frequency range may be ascribed to (i) the tail of the Drude-type absorption broadened at the high-density excitation, and (ii) the limited phase space for the low-energy interband transition in the 3D quasi-linear dispersion. Stimulated emission may be realized at much higher frequency in NIR and MIR as reported in graphene by measuring transient NIR reflectance[17]. We have also conducted the similar pump-probe experiment with a longer pump wavelength of 4 μm (see SI Note 8), which results in larger bleaching with a suppressed tail of the Drude absorption than that in the 2-μm pump. A smaller pump photon energy would suppress the scattering between hot electrons, which may be effective to achieve the stimulated emission. Among similar systems with massless-like Kane electrons, the multiterahertz stimulated emission has been achieved in a narrow-gap semiconductor $Hg_{1-x}Cd_xTe$ with quantum well structure at cryogenic temperature[42,43]. For $Cd_3As_2$, reducing the film thickness can change the band structure to a gapped topological insulator[44]. Such a band structure engineering with reduced temperature may enhance the possibility of realizing the light amplification because of increased density of states and suppressed recombination.

Comparing Figs. 2(a) and 2(b), the sign of $\Delta T/T$ does not always coincide with the sign of $-\Delta\sigma_1$, i.e., the transmittance can increase even if the absorption also increases. This can be explained by decrease of reflection. Figures 4(a)-4(d) show the change in transmittance $\Delta T/T$, the real-part dielectric function $\epsilon_1$, the refractive index $n$, and the extinction coefficient $\kappa$, respectively, for various delays with the pump fluence of 50 μJ/cm². The black curves show the data before the pump, and it changes to the red, and relaxes to the green and blue curves in time. The photoinjection of carriers elevates the screened plasma frequency at $\epsilon_1=0$ from 10 THz to a few tens of THz as shown in Fig. 4(b). Accordingly, $\epsilon_1$ and the refractive index $n$ in the multiterahertz regime significantly decreases, which suppresses the reflection loss at the surface.

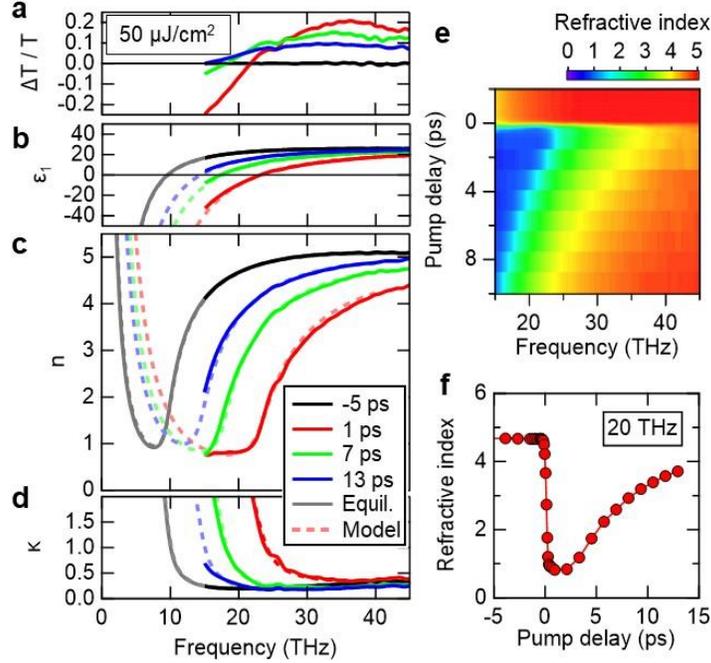

**Figure 4.** Change of response functions. (a)-(d) Transient transmittance $\Delta T/T(\omega)$, real-part dielectric function $\epsilon_1(\omega)$, refractive index $n(\omega)$, and extinction coefficient $\kappa(\omega)$, respectively, for various delays with the pump fluence of 50 μJ/cm². The broken curves show fitted results assuming the Drude response and interband transition. (e) 2D plot of the refractive index as a function of frequency (horizontal) and the pump delay (vertical). (f) Temporal evolution of refractive index at 20 THz as a function of the pump delay.

Our results demonstrating the large decrease of $n$ is in line with the interpretation of the recent ultrafast reflectivity measurement[22] probed at 2.8-4.4 μm (70-110 THz). Note that the analysis in Ref. 22 assumes that the extinction coefficient $\kappa$ is proportional to the absorption. Strictly speaking, however, $\kappa$ is defined as the spatial decay rate of the field due to both absorption *and* reflection and therefore is equal to $\text{sgn}(\epsilon_2)\sqrt{\sqrt{\epsilon_1^2+\epsilon_2^2}-\epsilon_1}/\sqrt{2}$. Since the permittivity $\epsilon_1 \sim 17$ and the absorption $\epsilon_2 \sim 4$ are comparable at MIR in $Cd_3As_2$[31] and they can be significantly modified by photoexcitation, the assumption in Ref. 22 may not be necessarily justified. Our time-domain spectroscopy directly provides spectral information of $\epsilon_1$ and $\epsilon_2$ $(=\sigma_1/\epsilon_0\omega)$, and thus quantitatively corroborates the reduction of $n$.

Figure 4(e) shows the 2D plot of $n$ as a function of frequency and pump delay at the pump fluence of 50 μJ/cm$^2$. Figure 4(f) shows the dynamics of $n$ at 20 THz, showing a drastic reduction from 4.7 to 0.80. Since $n$ is larger than $\kappa$ and the penetration depth in this frequency is as large as several μm in this multiterahertz frequency scale, the infrared response of thin films can be dominated by $n$ rather than $\kappa$. Therefore, the transient broadband small refractive index may be utilized for designing ultrafast optical switches. Recovery time of the reduced refractive index is determined by the carrier lifetime, which is ~10 ps in Cd$_3$As$_2$ and can be further shortened to 1 ps by doping for high-speed response[9]. Note that such a large reduction of $n$ itself around the plasma frequency can be expected in any photoexcited semiconductors. How much the refractive index is reduced and how broadly this occurs in frequency space depends on the scattering, because a short scattering rate elongates the tail of the Drude-type absorption and prevents $n$ from anomalous decrease. Although the scattering rate is enhanced in photoexcited Cd$_3$As$_2$ as it is in usual semiconductors, it tends to be robust against carrier density in topological materials with linear dispersion[45]. If one can use another topological semimetal with scattering time insensitive to the carrier density, it would realize unprecedented controllability of infrared response functions for novel optoelectronics and active metamaterials.

In summary, we conducted time-resolved spectroscopy for a photoexcited Cd$_3$As$_2$ thin film in the multiterahertz frequency region. Ultrafast probe of the broadband response functions clarified the dynamics of photoexcited carriers with thermalization and relaxation processes in Cd$_3$As$_2$. We also quantitatively demonstrated a large suppression of the refractive index by more than 80% due to the reduced permittivity and sharp Drude response. The results provide valuable information for designing infrared high-speed optoelectronic devices as well as for control of topological phase of matter by infrared light.

ASSOCIATED CONTENT

**Supporting Information**. The following files are available free of charge.

Sample fabrication, two-temperature model analysis, measurement of linear response function, pump-probe spectroscopy, correction in pump-probe spectroscopy analysis, double exponential fit of the carrier relaxation dynamics, weaker pump fluence, stronger pump fluence and effect of multi-photon absorption in the substrate, pump wavelength dependence. (PDF)

**Author Contributions**

R.M. conceived this project. M.G. fabricated the sample and S. S.-R. performed transmission electron microscopy studies, with guidance by S.S. N.K. and T.M. evaluated the linear response function. N.K. developed the pump-probe spectroscopy system, performed the experiments, and analyzed the data with help of Y.M., J.Y, and R.M. T.M. performed the TTM analysis. All the authors discussed the results. R.M. and N.K. prepared the manuscript with substantial feedbacks from all the coauthors.

**Notes**

ACKNOWLEDGMENT

   The authors thank Alexander Lygo for help with the sample characterization. This work was supported by JST PRESTO (Grant Nos. JPMJPR20LA and JPMJPR2006), JST CREST (Grant No. JPMJCR20R4), and in part by JSPS KAKENHI (Grants Nos. JP19H01817 and JP20J01422, and JP20H00343). R.M. also acknowledges partial support by Attosecond lasers for next frontiers in science and technology (ATTO) in Quantum Leap Flagship Program (MEXT Q-